\documentclass[aps,prl,showpacs,twocolumn]{revtex4}
\usepackage{bbm}
\usepackage{mathrsfs}
\usepackage{epsfig,psfrag}
\usepackage{amsmath,amsfonts,amssymb}
\usepackage[usenames]{color}

\begin{document}

\preprint{draft}

\title{Conformal Curves on $WO_3$ Surface }
\author{  A. A. Saberi, M. A. Rajabpour, S. Rouhani }
\address
{ Department of Physics, Sharif University of Technology, P.O. Box
11155-9161, Tehran, Iran }
\date{\today}

\begin{abstract}
We have studied the iso-height lines on the $\mathrm{WO_3}$ surface
as a physical candidate for conformally invariant curves. We have
shown that these lines are conformally invariant with the same
statistics of domain walls in the critical Ising model. They belong
to the family of conformal invariant curves called Schramm-Loewner
evolution (or $SLE_{\kappa}$), with diffusivity of $\kappa \sim 3$.
This can be regarded as the first experimental observation of SLE
curves. We have also argued that Ballistic Deposition (BD) can serve
as a growth model giving rise to contours with similar statistics at
large scales.
\end{abstract}

\pacs{68.37.-d, 05.45.Df}

\maketitle

The study of random surfaces, especially their statistical
properties as well as growth and evolution dynamics, has been
growing over the last two decades. They describe many important
problems of real surfaces appearing in condensed matter physics such
as deposited metal. In addition, this kind of problems is closely
related to other problems in physics such as fractured surfaces,
string theory, phase transitions in two dimensions etc
\cite{kondev99,stanley}. The aim of this paper is to show that
conformally invariant curves appear on the iso-height contours of
deposited films of $WO_3$. This is the first observation of SLE in
real physical system. We believe that the same result may be
observed for other surfaces.

Metal oxides are a large family of materials possessing various
interesting properties. One of the most interesting metal oxides is
tungsten oxide, $WO_3$, which has been investigated extensively
because of its distinctive applications such as
 electrochromic
\cite{Granqvist,Bueno,Azimirad,Kuai,Takeda}, photochromic
\cite{Avellaneda}, gas sensor \cite{Kim,György,Kawasaki},
photo-catalyst \cite{Gondal}, and photoluminescence properties
\cite{Feng}.\\ Many properties of tungsten oxide are related to its
surface structure (e.g. porosity, surface-to-volume ratio) and
surface morphology and statistics such as grain size and height
distribution of the sample. These properties are also affected by
conditions during the growth process such as deposition method. One
can change the statistics of the growth surface by imposing external
parameters such as annealing temperature which can even cause phase
transition in the sample \cite{saberi}.

On the other hand, calculation of various geometrical exponents of a
random Gaussian surface is of interest to theoretical physics. From
this point of view, we want to study the morphology and geometrical
statistics of experimental grown surfaces of $WO_3$. We consider
contour lines (the nonintersecting iso-hight lines) on the $WO_3$
samples and show that they are conformally invariant. This analysis
when applied to the contour lines of the ballistic disposition (BD)
model shows similar conformally invariant surface in large scales.
%%%%%%%%%%%%%%%%%%%%%%%%%%%%%%%%%%%%%%%%%%%%%%%%%%%%%%%%%%%%%%%%%%%%%%%%%%%%%%%%%%%%%%%%%%%%%%%%%%%%%%%%%%%%%%%%
\begin{figure}\begin{center}
\epsfxsize=7.5 truecm\center\epsfbox{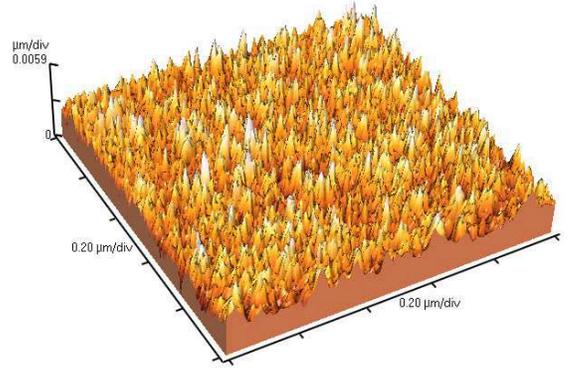}
 \narrowtext \caption{\label{Fig1}(color
online). AFM image of "as deposited" sample of $WO_3$
 thin film in a scale of 1$\mu$m $\times$1 $\mu$m with the resolution of 1/256 $\mu$m.}\end{center}
\end{figure}
%%%%%%%%%%%%%%%%%%%%%%%%%%%%%%%%%%%%%%%%%%%%%%%%%%%%%%%%%%%%%%%%%%%%%%%%%%%%%%%%%%%%%%%%%%%%%%%%%%%%%%%%%%%%%%%%%%%%
The scaling behavior of contour lines in the growth models are
widely investigated. For example in \cite{kondev} the authors
derived some of the universal exponents of contour lines, especially
their fractal dimension $D_0$ and its relation to the roughness
exponent $\alpha$. Moreover, Schramm and Sheffield recently showed
that the contour lines in a two-dimensional discrete Gaussian free
field are conformally invariant and belong to a wide class of
conformally invariant curves called Schramm-Loewner evolution
($SLE_{\kappa}$ ) curves, whose diffusivity $\kappa$ converges to 4
\cite{Schramm-Sheffield}-- every conformal curve in two dimensional
plane can be studied by SLE \cite{schramm} (to review about SLE see
\cite{cardy}). Moreover, it has been proved by Smirnov that some
domain walls in the critical Ising model, in the scaling limit, can
be described by $SLE$
\cite{Smirnov}.\\
In addition some physical systems have been studied recently using
some numerical methods. It has been shown that the statistics of
zero-vorticity lines in the inverse cascade of 2D Navier-Stokes
turbulence displays conformal invariance and can be described by
$SLE_6$ which is in the universality class of percolation
\cite{bernard1}. Similar studies have been done for inverse cascade
of surface quasigeostrophic turbulence \cite{bernard2}, domain walls
of spin glasses \cite{spin glass} and the nodal lines of random wave
functions \cite{Keatin}.

To study the geometrical statistics of the surface of $WO_3$, 34
samples were independently deposited on glass microscope slides with
the area $2.5
$$cm\times2.5$$cm$ using a thermal evaporation method in same
conditions. The deposition system was evacuated to a base pressure
of $\sim 4\times 10^{-3}$ Pa. The thickness of the deposited films
was considered to be about $200$ nm, measured by the stylus and
optical techniques.
%%%%%%%%%%%%%%%%%%%%%%%%%%%%%%%%%%%%%%%%%%%%%%%%%%%%%%%%%%%%%%%%%%%%%%%%%%%%%%%%%%%%%%%%%%%%%%%%%%%%%%%%%%%%%%%%%%%%%%
\begin{figure}\begin{center}
\epsfxsize=5.
truecm\center\epsfbox{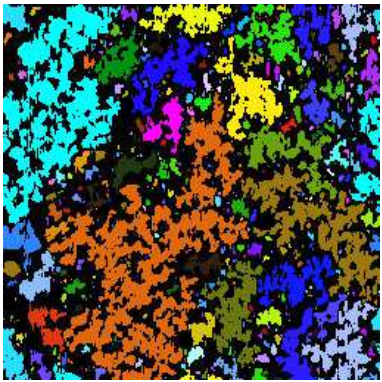}\caption{\label{Fig2}(color online).
The connected domains with positive heights on the $WO_3$ surface.
Negative height regions are black.}\end{center}
\end{figure}
%%%%%%%%%%%%%%%%%%%%%%%%%%%%%%%%%%%%%%%%%%%%%%%%%%%%%%%%%%%%%%%%%%%%%%%%%%%%%%%%%%%%%%%%%%%%%%%%%%%%%%%%%%%%%%%%%%%%
Using the atomic force microscopy (AFM) technics we have obtained
the height profiles $h(\textbf{r})$ of the rough surfaces with the
resolution of $1/256$ $\mu m$ in the scale of $1\mu m\times1\mu m$.
The AFM scans in this scale were performed in the various
non-overlapping domains (10 profiles for each sample) from the
centric region of the deposited samples (Fig. \ref{Fig1})--to ensure
that the boundary effects are negligible for the AFM profiles.\\
First we have determined the clusters by considering the connected
domains with like-sign heights (mean height is set to zero for a
range of heights around the mean value much less than
$h_{rms}$)(Fig. \ref{Fig2}), then the cluster boundaries (contour
lines) have been specified (Fig. \ref{Fig3}). We have obtained an
ensemble of such contour lines using MATLAB algorithm for contour
plot. The calculation of the fractal dimension of these contour
lines yields $D_0=1.38\pm0.02$ using the box-counting method. The
roughness exponent for the samples is calculated by using the
second-order structure function analysis, which gives
$\alpha=0.33\pm0.03$.

In order to investigate the scale dependency of the contours we have
analyzed their multifractal behavior.  The generalized fractal
dimensions are defined by $D_{q}=\frac{1}{q-1}lim_{a\rightarrow
0}\frac{\ln Z_{q}(a)}{\ln a}$ where $Z_{q}(a)=\sum\rho_{i}(a)^{q}$,
and the sum runs over all the non-overlapping boxes with size $a$
which cover the curve. $\rho_{i}(a)$ is the mass of the curve in the
$\emph{i}$th box. The multifractal spectrum of the contours is
independent of $q$ with the value of $1.38\pm0.03$. It shows that
the curves are scale invariant.
%%%%%%%%%%%%%%%%%%%%%%%%%%%%%%%%%%%%%%%%%%%%%%%%%%%%%%%%%%%%%%%%%%%%%%%%%%%%%%%%%%%%%%%%%%%%%%%%%%%%%%%
\begin{figure}\begin{center}
\epsfxsize=5. truecm\center\epsfbox{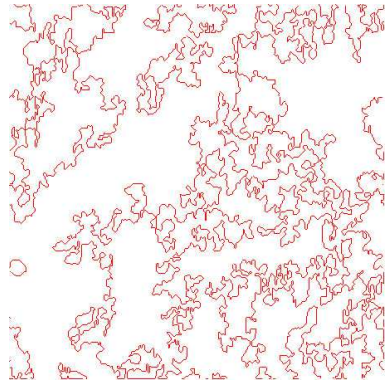}
\caption{\label{Fig3}(color online). Some of the positive height
cluster boundaries on a $WO_3$ surface (see also Fig.
\ref{Fig2}).}\end{center}
\end{figure}
%%%%%%%%%%%%%%%%%%%%%%%%%%%%%%%%%%%%%%%%%%%%%%%%%%%%%%%%%%%%%%%%%%%%%%%%%%%%%%%%%%%%%%%%%%%%%%%%%%%%%%%
To investigate more the scale dependency of contour lines, one can
check some of the exponents associated with clusters and loops. Fig.
\ref{Fig4}a shows the scaling of the average area of clusters $M$
with their radius of gyration $R$, $M\sim R^D$ which yields the
fractal dimmension of clusters $D=2-(8-\kappa)(3\kappa-8)/32\kappa$
\cite{Duplantier}. Fractal dimension of loops relates their length
$s$ to radius of gyration $R$ as $s\sim R^{D_{0}}$. In our case, it
is in good agreement with both the fractal dimension obtained using
the box-counting method and that of domain walls of critical Ising
model (Fig. \ref{Fig4}b). We have also checked the other scaling
relations, the average number of loops with radius $R$, $n(R)\sim
R^{-2+\alpha}$ (Fig. \ref{Fig4}c) and that with perimeter $s$,
$n(s)\sim s^{(-2+\alpha)/D_0}$ where $\alpha$ is the roughness
exponent (Fig. \ref{Fig4}d). Another is the scaling relation for the
number of loops with the area greater than $A$, $n(A)\sim
A^{-1+\alpha/2}$.

Consistency of the results for the contour lines of $WO_3$ samples
and Ising cluster boundaries suggests that a contour line of these
two belongs to the same universality class. However assumption of
translational and rotational symmetry within contour ensembles has
been made, which seems right. Therefore the iso-height contours of
$WO_3$ may be conformally invariant and belong to the $SLE_\kappa$
curves with $\kappa=3$. The fractal dimension $D_{0}=1+\kappa/8$ is
also consistent with our finding provided $\kappa=3$.

To examine this hypothesis directly, we first consider an arbitrary
placed horizontal line representing the real axis in the complex
plane across the AFM profiles. Then we cut the portion of each curve
$\gamma_{t}$ above the real line as it is in the upper half plane
$\mathbb{H}$.

Now let us to consider the chordal SLE in the $\mathbb{H}$ from $0$
to $\infty$ which is described by the Loewner equation
$\partial_{t}g_{t}(z)=\frac{2}{g_{t}(z)-\xi_{t}}$. If we
parameterize the contour lines in the $\mathbb{H}$ with the
dimensionless parameter $t$, $g_{t}(z)$ maps the upper half plane
except the curve up to time $t$, to $\mathbb{H}$ itself. For any
self avoiding curve $\gamma_{t}$ the driving function $\xi_{t}$ is a
continuous real function with the correspondence of
$g_{t}(\gamma_{t}(z))=\xi_{t}$. In the other words, $g_{t}(z)$ in
each time, maps the tip of the curve to a real point $\xi_{t}$.
\begin{figure}\begin{center}
\epsfxsize=8 truecm\epsfysize=8.8 truecm\center\epsfbox{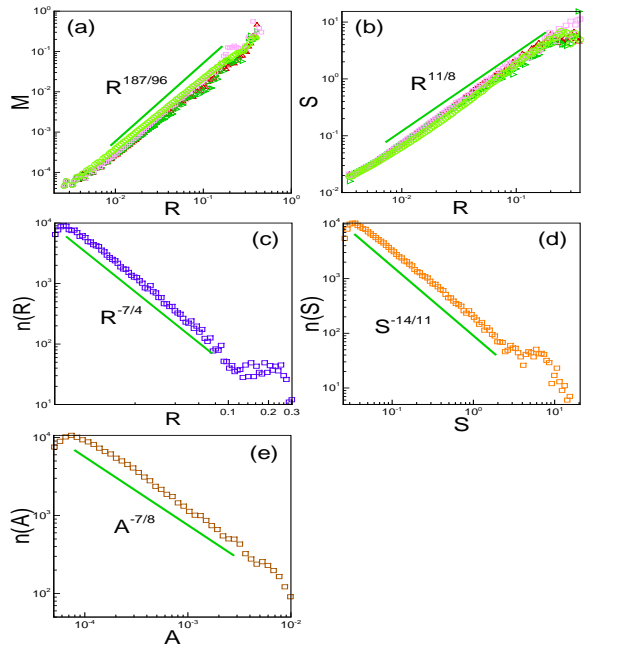}
 \narrowtext\caption{\label{Fig4}(color
online). Cluster and loop statistics for iso-height
 lines of $WO_3$ samples in scales 1$\mu$m $\times$1 $\mu$m (Squares)
 and 5$\mu$m $\times$5 $\mu$m, 10$\mu$m $\times$10 $\mu$m and 20$\mu$m $\times$20 $\mu$m.
  The size of the samples are rescaled to be 1
 with $256^2$ lattice points.
 (a) The average area $M$ of clusters versus the radius of gyration
 $R$.
 (b) The perimeter of loops $s$ versus the radius of gyration $R$.
 (c) The average number of loops with the radius of $R$.
 (d) The average number of loops with the perimeter $s$.
 (e) Number of loops of area greater than $A$. The error bars are
 almost the same size as the symbols in the scaling regions, and have not been drawn.
 These results are very close to the corresponding statistics
 of domain walls in the critical Ising model (the solid lines). Note that
 in Figs. (c), (d) and (e) the change in the exponents relative to the critical Ising model
 is due to the roughness exponent \cite{kondev}.}\end{center}
\end{figure}
Random curves whose driving functions are proportional to the
Brownian walk $B_{t}$ are conformally invariant. So, to investigate
the conformal invariancy of the contour lines we should explicitly
show that their driving function $\xi_{t}=\sqrt{\kappa} B_{t}$ i.e.,
its increments are identically and independently distributed and
$\langle\xi_{t}^2\rangle=\kappa t$, where $\kappa$ is diffusion
constant.

\begin{figure}\begin{center}
\epsfxsize=7.5truecm \epsfysize=5. truecm\center\epsfbox{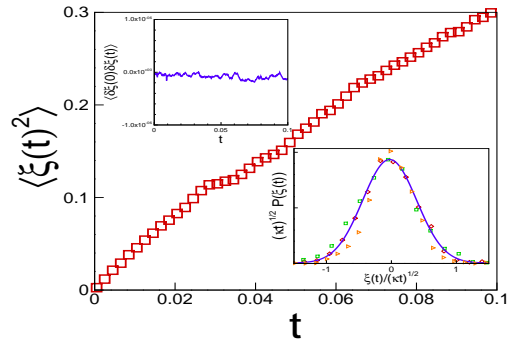}
\caption{\label{Fig5}(color online). Statistics of the driving
function $\xi(t)$. main frame: the linear behaviour of
$\langle\xi(t)^2\rangle$ with the slope of $\kappa=3\pm0.2$.
Lower-right inset: The probability density of $\xi(t)$ rescaled by
its variance $\kappa$ at different times $t=0.04, 0.06, 0.08$ is
Gaussian. Upper-left inset: The correlation function of the
increments of the driving function $\delta\xi(t)$.}\end{center}
\end{figure}

There are many numerical algorithms for extracting the driving
function of a given random curve by inversion of loewner equation
\cite{Kennedy}. We have used the successive discrete, conformal slit
maps -- based on the piecewise constant approximation of the driving
function --  that swallow one segment of the curve at each time step
\cite{Keatin}. To do this at first, we have determined all of the
contour lines $\gamma_{t}$s of the $WO_3$ samples as described above
by sequences of the points $\{z_0,z_1,\cdot\cdot\cdot,z_N\}$. we
obtained $2319$ such curves with the average number of points about
$1210$. After that all of the contour lines were mapped by
$\varphi(z)=z_Nz/(z_N-z)$. To avoid numerical errors only the
part of the curves corresponding to capacity $t\leq 0.25$ were used.\\
Then, by setting the starting point $z_0$ in $(0,0)$, all of the
points except the first one have been mapped with
$g_t(z)=\sqrt{(z-\xi_t)^2+4t}+\xi_t$ with $t=\frac{1}{4}Im(z_1)^2$
and $\xi_t=Re(z_1)$ and the resulting points have been renumbered by
one element shorter. The algorithm is repeated until the whole curve
is transformed. The final output of the process is a series of
driving functions obtained from the contour ensemble of the $WO_3$
samples. (The accuracy of the code has been checked on an ensemble
of SLE traces with known $\kappa$ where it yielded the correct value
with an error of $\sim3\%$).\\
As shown in Fig. \ref{Fig5}, the statistics of the driving function
--considering finite size effects-- converges to a Gaussian process
with variance $\langle\xi(t)^2\rangle=\kappa t$ and $\kappa=3\pm0.2$
and uncorrelated increments. It confirms that the contour lines of
the $WO_3$ samples are locally $SLE_3$ curves i.e., conformally
invariant.
%%%%%%%%%%%%%%%%%%%%%%%%%%%%%%%%%%%%%%%%%%%%%%%%%%%%%%%%%%%%%%%%%%%%%%%%%%%%%%%%%%%%%%%%%%%%%%%%%%%%%%%%%%%%%
To investigate the effect of the scale of AFM images on the
statistics of the contour lines, we took over $10^2$ AFM profiles
from samples at each of three scales 5$\mu$m$\times$5 $\mu$m,
10$\mu$m$\times$10 $\mu$m and 20$\mu$m$\times$20 $\mu$m with a fixed
number of lattice points $256^2$. After rescaling of size to one, we
observe no conclusive differences within numerical errors (Fig.
\ref{Fig4}).

\begin{figure}\begin{center}
\epsfxsize=7.truecm \epsfysize=7.8truecm \center\epsfbox{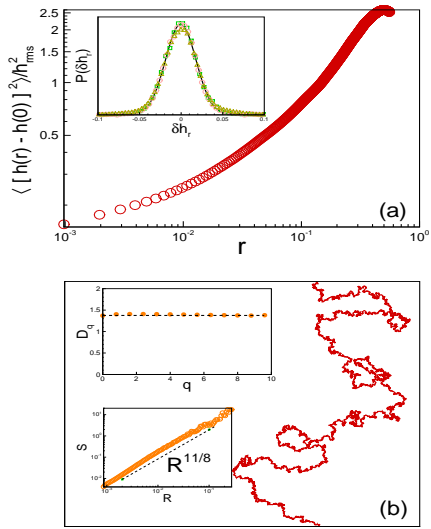}
\caption{\label{Fig6}(color online). (a) Second-order structure
function of height differences and, in the inset, rescaled and
normalized probability density functions for r=0.02, 0.04 and 0.06
compared to a Gaussian density (solid line). Data have been obtained
from a simulated BD growth surface in square domain of size 1 with
$1024^2$ lattice points. (b) An instant contour line and, in the
upper-inset, generalized multifractal dimensions for the BD contour
lines which are independent of q within statistical errors. The
dashed line is set in the $11/8$. Lower-inset shows the scaling of
the perimeter of the loops $s$ versus the radius of gyration $R$
compared with the one in the Ising model (dashed line).}\end{center}
\end{figure}

In the following, due to some similarities between some of the
results for $WO_3$ surface and BD model, we wish to briefly discuss
this model.\\
To investigate the behavior of the contour lines of BD growth
surface, we simulated $10^2$ independent samples with the size of
$1024\times1024$. The roughness exponent of the samples is
$0.36\pm0.02$ \footnote{The exact value of the roughness exponent of
BD model in two dimensions is not known so far \cite{Reis1}.}. As
shown in Fig. \ref{Fig6}a, the statistics of the heights are
self-similar and the height increments have approximately Gaussian
statistics \cite{Reis2}.\\
The multifractal spectrum of the BD contour lines has been
numerically obtained which shows that the contour lines seem to be
scale invariant. The fractal dimension of the lines is $1.39\pm0.03$
(Fig. \ref{Fig6}b).
In the first view, these results show that the BD's contour lines
may belong statistically to the same universality class as the
domain walls in the critical Ising model. Applying the Inverse
loewner equation, we have analyzed the driving function of the
iso-height curves. This analysis shows that in some large scales the
increments of the driving function are independently distributed
with Gaussian statistics of variance $\kappa=3.1\pm0.2$.

A point which remains is the difference between the standard way of
generating an interface in the Ising model and the one considered in
this paper. To generate chordal SLE in the half plane for the Ising
model, one usually takes fixed boundary conditions on either side of
the origin. Whereas in the procedure proposed here, we choose an
arbitrary horizontal line which clearly has more complex boundary
conditions on it. Certainly different boundary conditions can
introduce different drift terms in the driving function. Question is
which boundary condition with the above procedure leads to chordal
SLE in the Ising model. We have run simulations on the two
dimensional critical Ising model with periodic boundary conditions
on the square lattice of size $1024^2$, and found that the resulting
driving function has leading term (considering finite size effects)
with the same statistics of Brownian motion with diffusivity of
$\kappa=2.9\pm 0.2$. Details will appear in a forthcoming work.

In conclusion, we have found an experimental example, the iso-height
lines of the $WO_3$ surface, for SLE curves. It has been numerically
shown that the statistics of the contour lines are conformally
invariant and it is very similar to the statistics of the domain
walls in the critical Ising model. However, this approach can be
used for other experimental samples to know more about the relation
of the statistics of the contour lines to the various geometrical
and physical properties of the surfaces. Moreover, we have proposed
a model of growth , ballistic deposition model whose contour lines
seem to belong to the same universality class as the Ising model at
large scales.

We wish to thank A.Z. Moshfegh, R. Azimirad and S. Vaseghinia for
providing AFM data. We would like to thank B. Fallah Hassanabadi,
S.M. Fazeli, R. Sepehrinia and especially R. Nourafkan and M. D.
Niry for their useful discussions. We also wish to thank F. D. A.
Aar\~{a}o Reis, J. Cardy, O. Schramm and R. Stinchcombe for
providing helpful comments on this manuscript.

\end{document}